%% file: main.tex
\documentclass[runningheads]{llncs}
\usepackage{graphicx}
\usepackage{marvosym}
\usepackage{multirow}
\usepackage{lscape}
\usepackage{booktabs}
\usepackage{tabularx}
\usepackage{makecell}
\usepackage{amsmath}
\usepackage{diagbox}
\usepackage{lipsum}

\begin{document}

\title{A Context Model for Personal Data Streams\thanks{Xiaoyue receives funding from the China Scholarships Council (No.202107820014). Marcelo, Fausto and Matteo receive funding from the project “DELPhi - DiscovEring Life Patterns” funded by the MIUR (PRIN) 2017.}}

%
\author{Fausto Giunchiglia\orcidID{0000-0002-5903-6150} \and
Xiaoyue Li\textsuperscript{\Letter}\orcidID{0000-0002-0100-0016} \and
Matteo Busso\orcidID{0000-0002-3788-0203} \and Marcelo Rodas-Britez\orcidID{0000-0002-7607-7587} }

%
\institute{ 
Department of Information Engineering and Computer Science, University of Trento, Trento, Italy\\
\email{\{fausto.giunchiglia, xiaoyue.li, matteo.busso, marcelo.rodasbritez\}@unitn.it}}
\maketitle              

\begin{abstract}
\input{00Abstract}
\end{abstract}

\input{01Introduction}

\input{02TheSituationalContext}

\input{04Schema}

\input{05CaseStudy}

\input{07Conclusion}

\bibliographystyle{splncs04.bst}
\bibliography{bibliography.bib}

\end{document}

%% file: 00Abstract.tex
We propose a model of the \textit{situational context} of a person and show how it can be used to organize and, consequently, reason about massive streams of sensor data and annotations, as they can be collected from mobile devices, e.g. smartphones, smartwatches or fitness trackers. 

The proposed model is validated on a very large dataset about the everyday life of one hundred and fifty-eight people over four weeks, twenty-four hours a day.
\keywords{Personal Situational Context \and Data Streams.}

%% file: 01Introduction.tex
\section{Introduction}
\label{01-Introduction}

A lot of prior work has focused on collecting and exploiting massive streams of data, e.g., sensor data and annotations. A first line of work has concentrated on 
using the streams of personal data for learning daily human behavior, including physical activity, see, e.g., \cite{patterson2021smartphone}, assessment personality states, see, e.g., \cite{ruegger2020personality}, and visiting points of interest \cite{do2013places}.
The \textit{Reality Mining} project \cite{eagle2006reality} collected smartphone sensors, including call records, cellular tower IDs, and Bluetooth proximity logs to study students' social networks and daily activities. In the same vein, the \textit{StudentLife} project \cite{wang2014studentlife,harari2020sensing} employed smartphone sensors and questionnaires as the means for inferring the mental health, academic performance, and other behavioral trends of university students, under different workloads and term progress. Slightly different in focus, but still based on the collection of streams of data, is the work on the Experience Sampling Method (ESM). The ESM is an intensive longitudinal social and psychological research methodology, where participants are asked to report their thoughts and behaviours \cite{van2017experience}. Here the focus is not so much on learning from the sensor data but, rather, on collecting the user provided answers.
In all this work, little attention has been posed on how to represent and manage these data streams. The most common solution has been that of collecting these data, \textit{as is}, into (multiple) files in some common format, e.g. CSV. Which was good enough, given that data were exploited \textit{a posteriori}, once the data collection was finished, by doing the proper off-line data analysis. 

Our focus is on the exploitation of data, \textit{at run-time}, while being collected, as the basis for supporting person-centric services, e.g., predicting human habits or better human-machine interaction. This type of services are in fact core for the development of \textit{human-in-the-loop} Artificial Intelligence systems \cite{KD-2022-Bontempelli-lifelong}. Towards this end, our proposed solution is to represent the input streams, no matter whether coming from sensors or from the user feedback, as sequences of \textit{personal situational contexts} \cite{KD-2021-Giunchiglia-EtA-MRCl}. Here by context we mean \textit{``a theory of the world which encodes an individual’s subjective perspective about it”} \cite{KD-1993-giunchiglia,P-2017-Giunchiglia.a}.
Many challenges still need to be solved towards this goal. For instance, these data are highly heterogeneous, e.g., categorical, numerical, in natural language, and unstructured, usually collected with different time frequencies. Furthermore, different data may be at different levels of abstraction, for instance the current location can be described as, e.g., GPS coordinates, my office, the University, or the city of Trento.  

The main goal of this paper is to provide a representation of data streams at the \textit{knowledge level} \cite{newell1982knowledge}, rather than only at the \textit{sensor or data level}, fully understandable by the user, in the user terms, thus enabling the kind of Human-Machine interactions which we need. 
We realize this requirement by representing streams as sequences of situational contexts, and by modeling them as Knowledge Graphs (KGs) \cite{bonatti2019knowledge}. In this context, by KG we mean a graph where the nodes are the entities involved in the current user context, e.g., friends, the current location, the current event, for instance a meeting, while links are the relations occurring among entities, e.g., the fact that two people are classmates or that a person is on a car or talking to another person. Notice how various notions of context model have been proposed in the past. Some work focused on representing the current situation with reference to the location, see, e.g., \cite{P-1994-Schilit}. Other approaches have used hierarchical context models \cite{wang2004hk}.  
However, these proposals did not deal with the problem of how to provide an abstract user-level representation of ever growing streams of data.

The proposed design the knowledge level representation of the personal situational context is articulated in three steps, as follows:
\begin{enumerate}
    \item An abstract conceptualization of the notion of context in terms of the person space and time localisation plus the people and objects populating the context itself;
    \item A schema of the KG,  what we call an ETG (Entity Type Graph), which defines the data structure used the current situational context as it occurs in a certain period of time; 
    \item The actual data streams, memorised as sequences of context KGs each with the same ETG, differently populated.
\end{enumerate}

\noindent 
%
The paper is organized as follows. Section \ref{02TheSituationalContext} formalizes the notion of situational context. 
Section \ref{04-ERmodel} described the details of the situational context KG. Section \ref{05CaseStudy} presents a large scale case study.  Finally, in Section \ref{07-Conclusion}, we present our conclusions.

%% file: 02TheSituationalContext.tex
\section{The Situational Context}
\label{02TheSituationalContext}

A situational context represents a real world scenario from the perspective of a specific person, whom we call \texttt{me}, e.g., Mary. A \textit{Life sequence} is a set of situational contexts during a certain period of time. We define the life sequence of \texttt{me}, $S(me)$, as follows:

\begin{equation}
\setlength{\abovedisplayskip}{5pt}
\setlength{\belowdisplayskip}{5pt}
S\left(me\right) = \langle C_1\left(me\right), C_i\left(me\right), \dots, C_n\left(me\right) \rangle; \ \ \ 1 \leq i \leq n
\label{eq:S-C}
\end{equation} 

\noindent 
where $C_i$ is the $i_{th}$ situational context of \texttt{me}. We assume that \texttt{me} can be in only one context at any given time, based on the fact that a person can be in only one location at any time. Hence, $S$ is a sequence of \texttt{me}'s contexts, occurring one after the other, strictly sequentially, with no time in between.
%
%
In turn, we model the \textit{Situational context} of \texttt{me} $C(me)$ as follows:

\begin{equation}
\setlength{\abovedisplayskip}{5pt}
\setlength{\belowdisplayskip}{5pt}
C(me) = \langle L(C(me)), E(L(C(me))) \rangle. 
\label{eq:C-L-E}
\end{equation} 

\noindent
In the following, we drop the argument \texttt{me} to simplify the notation. $L(C)$ is the (current) \textit{Location} of \texttt{me}. $L(C)$ defines the boundaries inside which the current scenario evolves. The location is an endurant, which is wholly present whenever it is present, and it persists in time while keeping its identity \cite{P-2017-Giunchiglia.a}. $E(L(C))$ is an \textit{Event} within which \texttt{me} is involved. The event is a perdurant, which is composed of temporal parts \cite{P-2017-Giunchiglia.a}. $L(C)$ and $E(L(C))$, as the priors of experience, define the scenario being modeled and the space-time volume within which the current scenario evolves. This is a consequence of the foundational modeling decision that contexts are the space-time prior to experience. In other words, the situational context of \texttt{me} is univocally defined by \texttt{me}'s \textit{spatial position} and \textit{temporal position}. In practice, any electronic device can easily provide us with the spatial position (via GPS, annotations, etc.) and temporal position (via timestamp) of a person. 

In a certain context, \texttt{me} can be inside one or multiple locations as follows:%
%
%
%

\begin{equation}
\setlength{\abovedisplayskip}{5pt}
\setlength{\belowdisplayskip}{5pt}
L(C) = \langle L_1(C), L_i(C),\dots , L_n(C) \rangle; \ \ \ 1 \leq i \leq n
\label{eq:L-subL}
\end{equation} 

\noindent
where $L_i(C)$ is a spatial part of $L(C)$, we call $L_i(C)$ is a \textit{sub-location} of $L(C)$. If \texttt{me} is inside one location, we have $ L(C)= L_1(C)=\dots = L_n(C)$, and the context is static, e.g., Mary is at the university library, or Mary is at home. Otherwise, the context is dynamic, e.g., Mary travels around Trento ($L(C)$), going from the university ($L_1(C)$), to the central station ($L_2(C)$), and then to her home ($L_3(C)$). 
%
%
%
Inside contexts, multiple events will occur:

\begin{equation}
\setlength{\abovedisplayskip}{5pt}
\setlength{\belowdisplayskip}{5pt}
E(L(C)) = \langle E_1(L(C)), E_i(L(C)), \dots , E_n(L(C)) \rangle; \ \ \  1 \leq i \leq n
\label{eq:E-subE}
 \end{equation}

\noindent 
where $E_i(L(C))$ is a part of $E(L(C))$. We call an $E_i(L(C))$ a \textit{sub-event} of $E(L(C))$. Different sub-events may occur in parallel or be sequential or mixed, but a sub-event can not be part of another sub-event. A simple event is the event where $E(L(C)) = E_1(L(C))=\dots = E_n(L(C))$. A complex event is the event where there are multiple distinct sub-events.

%
Finally, the context contains various types of things interacting with one another. We define a \textit{Parts of a Context} as follows:

\begin{equation}
\setlength{\abovedisplayskip}{5pt}
\setlength{\belowdisplayskip}{5pt}
P(C) = \langle me, \{P\}, \{O\}, \{F\}, \{A\} \rangle
\label{eq:P}
\end{equation} 

\noindent 
where $\{P\}$ and $\{O\}$ are, respectively, a set of \textit{persons} (e.g., Bob) and \textit{objects} (e.g., Mary's smartphone) populating the current context. $\{F\}$ and $\{A\}$ are, respectively, a set of \textit{functions} and \textit{actions} involving \texttt{me}, persons and objects.
%
We define a \textit{Generic object} $G$, consisting of \texttt{me}, $\{P\}$, and $\{O\}$, i.e., $G = \texttt{me} \cup \{P\} \cup \{O\}$. Functions define the roles that different generic objects have towards one another \cite{giunchiglia2017teleologies}. Thus a person can be a \textit{friend} with another person, a horse can be a \textit{transportation means} for person, while a phone can be a \textit{communication medium} among people. Functions are endurants.
Actions model how generic objects $G$ change in time \cite{giunchiglia2017teleologies}, e.g., Mary touches her smartphone in a certain moment, while she walks or eats at some other times. Actions are perdurants.
Functions are characterized by the set of actions which enable them \cite{giunchiglia2017teleologies}. Thus for instance, the function \textit{friend} might be associated with the actions \textit{talking to}, \textit{helping}, or \textit{listening to}. Similarly, a smartphone (i.e., $G_a$) can be recognized as an entertainment tool for Mary (i.e., $G_b$), because the smartphone allows certain actions related to the entertainment of Mary, e.g., playing videos, playing music, etc. Hence, for two generic objects $G_a$ and $G_b$, in the context, we have the following:

\begin{equation}
\setlength{\abovedisplayskip}{5pt}
\setlength{\belowdisplayskip}{5pt}
    F(G_a, G_b) =  \langle A_1(G_a, G_b), \dots , A_n(G_a, G_b) \rangle; \ \ \  
    \label{eq:F-A}
\end{equation}

\noindent 
where a function $F$ relates $G_a$ with $G_b$, namely, it is associated with the set of actions ($A_1, \dots, A_n$) involving $G_b$ that $G_a$ can do or allow.

%% file: 04Schema.tex
\begin{table}[!htp]
  \caption{Properties of the principal Entity types of a situational context.}\label{Propertytab}
  \resizebox{\textwidth}{25mm}{
    \begin{tabular}{|l|l|l|l|l|}
      \hline
      \diagbox[innerwidth=4.5cm]{Property types}{Entity types}&  Location; Sub-location & Event; Sub-event&  Person; \texttt{me} & Object \\
      \hline
      \makecell[l]{Spatial property: relating to or\\ occupying space}  &  \makecell[l]{Coordinates \\ Volume } & None & Coordinates &  Coordinates \\
      \hline
      \makecell[l]{Temporal property: relating\\ to time}  &  None & \makecell[l]{Start-EndTime} & None &  None \\
      \hline
      \makecell[l]{Function property: indicating\\ attributes of functions }  &  Location functions & None & Person functions &  Object functions \\
      \hline
      \makecell[l]{Action property: indicating\\ attributes of actions }  &  None & None & Person actions &  Object actions \\
      \hline
      \makecell[l]{External property: relating to\\ outward features}  &  \makecell[l]{Name\\ID} & \makecell[l]{Name\\ID} & \makecell[l]{Name\\ID\\Gender} &  \makecell[l]{Name\\ID\\Color} \\
      \hline
      \makecell[l]{Internal property: relating to\\ persons' internal states }  & None & None & \makecell[l]{InPain\\ InMood\\ InStress} &  None \\
      \hline
    \end{tabular}
  }
\end{table}

\section{The Entity Type Graph}
\label{04-ERmodel}


We define Location, Sub-location, Event, Sub-event and Generic object as \texttt{Entity types (etypes)}, where an entity is anything which has a name and can be distinctly identified via its properties and where, in turn, an etype is a set of entities. Functions and Actions are modeled as \texttt{Object properties} representing the relations among Generic objects. In Table \ref{Propertytab}, we define and provide examples of \texttt{Spatial}, \texttt{Temporal}, \texttt{External}, and \texttt{Internal data property types} as well as of \texttt{Function} and \texttt{Action object property types}.

\begin{figure}
\centering
\setlength{\abovecaptionskip}{0cm}
\includegraphics[width=\textwidth]{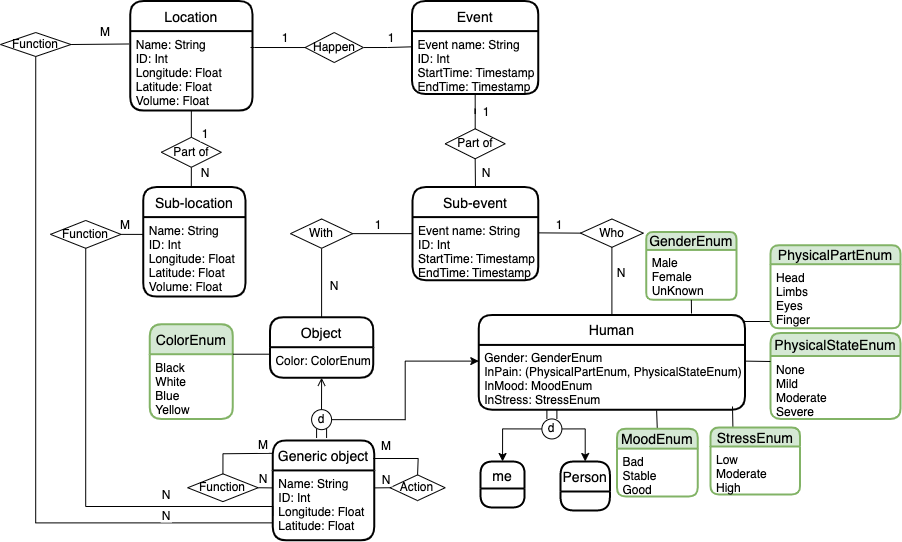}
\caption{An example of ETG modeling the situational context.} 
\label{fig:ETG}
\vspace{-0.5cm}
\end{figure}
\noindent
We represent the schema of the situational context of \texttt{me} as an eType Graph (ETG), i.e., an Enhanced Entity-Relationship (EER) model. See  Figure~\ref{fig:ETG} for a simplified version of an ETG representing a personal situational context. An ETG is a knowledge graph where nodes are etypes, decorated by data properties, which in turn are linked by object properties. 
%
Each etype (represented as a box) is decorated with its data properties. For example, the etype Human has the data property Gender, and the data type (the green box) of Gender is GenderEnum. 
Also, etypes are connected with object properties showing their relations (represented as rhombuses). One such example, is the relation \textit{With} which in turn is associated its own cardinality. Finally, 
as from EER models, it is possible to have inheritance relations among he etypes, e.g., a Generic Object is specialized into Object and Human. 

Given that a context is represented by a single ETG, we represent the evolution in time of the life of a person as a sequence of ETGs, each representing the state of affairs at a certain time and for a certain time interval. In turn, this sequence of ETGs is populated by the input data streams, where each element of the stream will populate the ETG for that time slot. Of course for each input stream there will be a dedicated suitable property for the proper etype. Thus, for instance, the GPS will populate the data property \textit{GPSLocation} of the etypes \textit{person} and/ or \textit{phone}, while the label of a location, e.g., \textit{Trento} will be used to create an object property link between the etype \textit{person} and the etype \textit{location}. Given the above, a life sequence, as defined in Section 2, is just a sequence of contexts satisfying a certain property, namely, a subset of the overall sequence of ETGs, populated by the input data streams. So for instance we may have Mary's life sequence of her moving around in Trento, see example above, or we can have the life sequence of all the times she has studied in her office at the University in the last year. Notice that this latter life sequence is composed of contexts which are not adjacent in time. This is a very powerful representational mechanism which can be used, for instance, to represent \textit{habits} as (not necessarily adjacent) life sequences occurring recursively with a certain frequency.

%% file: 05CaseStudy.tex
\section{Case Study}
\label{05CaseStudy}

To validate the formalization described above, we describe how it can be used to represent the Smart University stream dataset (SU).\footnote{See \url{https://livepeople.datascientia.eu/dataset/smartunitn2} for a detailed description of the dataset plus the possibility of downloading it.} The app used for the data collection is called \texttt{iLog} \cite{zeni2014ilog,2020-zeni1}. The SU data set has been  used in a large number of case studies, see, e.g., \cite{zeni2019fixing,zhang2021putting}.
%
 SU has been collected from one hundred and fifty-eight university students over a period of four weeks. It contains 139.239 annotations and approximately one terabyte of data. The dataset is organized into multiple datasets, one for each \texttt{me}, where each dataset is associated with a unique identifier across all types of data. The annotations done by each \texttt{me} are generated every half-hour based on the answers of the participants to four closed-ended questions.  Based on this, the best choice is to build a sequence of ETGs, one for every half an hour for each \texttt{me}. The four questions are ``Where are you?", ``What are you doing?", ``With whom are you?", and ``What is your mood?" and are based on the \textit{HETUS} (\textit{Harmonized European Time Use Surveys}) standard.\footnote{ https://ec.europa.eu/eurostat/web/time-use-survey.}

\begin{figure}[!htb]
\centering
\vspace{-0.1cm}
\setlength{\abovecaptionskip}{0cm}
\includegraphics[width=\textwidth]{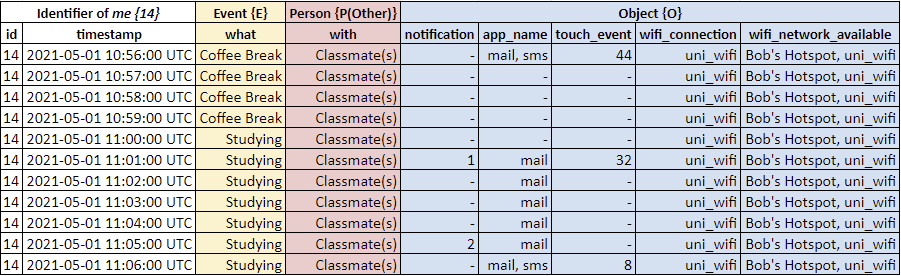}
\caption{The current event, the people and the object with Me.}
\label{fig:DF3}
\vspace{-0.6cm}
\end{figure}

Figures \ref{fig:DF2} and \ref{fig:DF3} provide a small, clean and anonymized subset of SU.  In both figures, the first part (in white) provides the timestamps when this data were collected. In the first figure, the location of me (in green) is represented together with some of her attributes (in orange). The second figure reports the current event (in yellow) in which me is involved, her function towards the person she is with (in red) and her phone with some of its attributes (in blue). 
It is easy to compare the contents of Figures \ref{fig:DF2} and \ref{fig:DF3} with the notions defined in the previous sections. Let us consider some examples:

\begin{figure}[!htb]
\centering
\vspace{-0.4cm}
\setlength{\abovecaptionskip}{0cm}
\includegraphics[width=\textwidth]{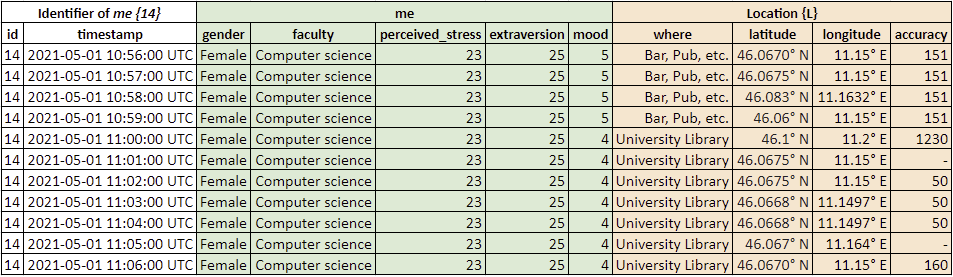}
\caption{Me and the current location.}
\label{fig:DF2}
\vspace{-0.2cm}
\end{figure}

\begin{itemize}
  \item \texttt{Human Entities}: They are \texttt{me} and \texttt{Person}, both associated with External and Internal properties. 
  \item \texttt{Human's External Properties}: They are mainly collected synchronously and are represented by the variables ``gender" and ``faculty". 
  \item \texttt{Human's Internal Properties}: They are both synchronic, i.e., ``extraversion", and diachronic, i.e.,``mood". 
  \item \texttt{Location Entity}: It is defined by  ``where" and it is annotated by the data properties ``latitude" and ``longitude" (with their respective ``accuracy"). 

\end{itemize}

\noindent
According to the research purpose, many additional data points may be used as proxies for characterizing the main notions of the context. For instance, concerning the \texttt{Location}, the WiFi router can be used as a proxy of a facility; a question posed in the online questionnaire about a daily routine can be a proxy of a travel path. By imputation on the GPS, it is possible to derive the Point Of Interest (POI), which can be understood as the set of \texttt{Objects} surrounding a given spatial coordinate. And so on.

%% file: 07Conclusion.tex
\section{Conclusion}
\label{07-Conclusion}

This paper proposes a  model of the situational context of a person and it shows how it can be used to provide a knowledge level representation over the data collected in time, both sensor data and user provided label, from mobile devices.